\documentclass[a4paper,fleqn,10pt]{article}

\usepackage{graphicx}
\usepackage{subfigure}
\usepackage{latexsym}
\usepackage{amsmath,amssymb}
\usepackage{dcolumn}

\begin{document}

\title{\bf Generalized holographic Ricci dark energy and generalized second law of thermodynamics in Bianchi type I universe}
\author{\normalsize En-Kun Li$^a$,
       Yu Zhang$^{a}$\thanks{Corresponding Author(Y. Zhang): Email: zhangyu\_128@126.com},
       Jin-Ling Geng$^a$ and Peng-Fei Duan$^b$ \\
   \normalsize \emph{$^a$Faculty of Science, Kunming University of Science and Technology, Kunming 650500, China}\\
   \normalsize \emph{$^b$City College, Kunming University of Science and Technology, Kunming 650051, China}}

\date{}
\maketitle

\begin{abstract}
  Generalized second law of thermodynamics in the Bianchi type I universe with the generalized holographic Ricci dark energy model is studied in this paper. The behavior of dark energy's equation of state parameter indicates that it is matter-like in the early time of the universe but phantom-like in the future. By analysing the evolution of the deviations of state parameter and the total pressure of the universe, we find that for an anisotropic Bianchi type I universe, it transits from a high anisotropy stage to a more homogeneous stage in the near past. Using the normal entropy given by Gibbs' law of thermodynamics, it is proved that the generalized second law of thermodynamics does not always satisfied throughout the history of the universe when we assume the universe is enclosed by the generalized Ricci scalar radius $R_{gr}$. It becomes invalid in the near past to the future, and the formation of the galaxies will be helpful in explaining such phenomenon, for that the galaxies's formation is an entropy increase process. The negative change rates of the horizon entropy and internal entropy occur in different period indicates that the influences of galaxies formation is wiped from internal to the universe's horizon.
\end{abstract}

\textbf{Keywords:} second law of thermodynamics; Bianchi type I universe; generalized holographic Ricci dark energy.

\textbf{PACS:} 98.80.-k, 95.36.+x

\section{Introduction}
\label{si}

Numerous observations such as Type Ia Supernovae, Cosmic Microwave Background Radiation, and Sloan Digital Sky Survey, provide quit precise evidences that our universe is undergoing an accelerated expansion phase \cite{Riess1998,Perlmutter1999,Spergel2003,Tegmark2004,Seljak2005,Spergel2007}. There are two representative ways to explain this phenomenon: one is introducing the dark energy (DE) with negative pressure in general relativity; another is modifying the gravity on the long distance \cite{Nojiri2007,Chen2008,Nojiri2011,Capozziello2011,Bamba2012}. Until now, there are a lot of DE models and modified gravity models have been put forward.

In this article, among various DE models, we concentrate on the holographic dark energy (HDE) model. The HDE is arising from the holographic principle \cite{Hooft,Susskind1995}, which links the energy density of DE to the cosmic horizon, attempting to examine the nature of DE in the framework of quantum gravity. The energy density of HDE is defined as
\begin{eqnarray}
  &\rho_h = 3 \mathcal{C}^2 M_p^2 L^{-2}, \label{ex}
\end{eqnarray}
where $M_p$ is the reduced Planck mass, $M_p^2 =(8\pi G)^{-1}$, $\mathcal{C}^2$ is a dimensionless model parameter, $L$ indicates the infrared (IR) cutoff radius \cite{Li2004,Hsu2004}. Based on this principle, researches on the HDE have attracted so many scientists, and a lot of remarkable works have been done in this field \cite{Setare2006,Setare2007A,Setare2007B,Setare2008,Setare2009,Cai2010,Karami2013}. In order to examine the HDE, one should give a special form of the IR cutoff. Until now, there are many choices have been taken as the IR cutoff radius, such as the Hubble radius \cite{Horava2000,Thomas2002}, the particle horizon \cite{Bousso1999,Fischler}, the future event horizon \cite{Huang2004}, the cosmological conformal time \cite{Cai2007,Wei2008}, the Ricci curvature scalar \cite{Gao2009}, or other generalized IR cutoff radius \cite{Granda2008,Granda2009,Chen2009}.
Among these HDE models, Xu et al. have considered generalized holographic and Ricci DE models \cite{Xu2009B}. In their work, the energy densities of DE are given as $\rho_h = 3 \alpha M_p^2 H^2 g(R/H^2)$ and $\rho_R = 3 \alpha M_p^2 R f(H^2/R)$, where $f(x)$ and $g(y)$ are functions of the variables $R/H^2$ and $H^2/R$. Nowadays, various works show that the HDE model is in fairly good agreement with the observational data \cite{Xu2009,Xu2010,Zhang2009,Micheletti2010,Wang2010,Duran2011,Radicella2010A}, and researches on different scenarios about the HDE can be found in Refs: \cite{Quartin2008,Karwan2008,Zhang2010,Setare2007C,Zhang2008,Feng2008,Chimento2012,Chimento2013,Chattopadhyay2013,Guo2007,Li2014,Li2014A}.

Thermodynamics behavior of the accelerated expansion universe driven by DE is one of the important questions in cosmology \cite{Wang2001,Frolov2003,Calcagni2005,Danielsson2005,Bousso2005,Cai2007D,Akbar2007,Sheykhi2007,Sheykhi2009,Myung2009,Abdolmaleki2014}. In different scenarios, Wang et al. \cite{Wang2006} find that the generalized second law of thermodynamics (GSLT) is valid on the apparent horizon, but invalid on the event horizon. Karami et al. \cite{Karami2010} find that in the non-flat universe, the validity of GSLT for the cosmological event horizon depends on the equation of state parameter (EoS) of DE. Bhattacharya et al. \cite{Bhattacharya2012} consider two types of DE models, which are raised by Xu et al. \cite{Xu2009B}, and find that neither the first law of thermodynamics nor the GSLT is valid on the horizon of IR cutoff radius $L$.
Moreover, Radicella and Pav\'{o}n proved that the united dark fluid model could fulfill the GSLT, but its soundness is in doubt for the entropy's first and second derivatives present a rather peculiar and sharp oscillation \cite{Radicella2014}.
More works about GSLT for the universe driven by HDE can be found in \cite{Radicella2010B,Radicella2012,Debnath,Sharif2011,Sharif2011A,Sharif2012,Khodam2012,Farajollahi2013,Padmanabhan2014,Sheykhi2014}.

Moreover, recent observation data from WMAP require that the universe should achieve a slightly anisotropic special geometry in spite of the inflation \cite{Hinshaw2007,Jaffe2006,Campanelli2007,Hoftuft2009}. Therefore, in an attempt to understand the observed small amount of anisotropy in the universe  better, the Bianchi type models have been studied by several authors \cite{Singh2010,Yadav2011A,Yadav2011B,Kumar2011,Amirhashchi2011,Singh2013,Singh2014}. Our aims are to investigate the evolution of the Bianchi type I (BI) universe with the HDE and interacting dark matter (DM), and the GSLT for the accelerated expanding BI universe driven by DE and DM. In Sect. \ref{sm}, we give a brief review of the anisotropic metric and its field equations. In Sect. \ref{sr}, the evolution of the energy densities and the EoS of DE are studied. In Sect. \ref{st}, the GSLT for the BI universe enclosed by the generalized Ricci scalar radius are studied. Sect. \ref{sc} is the conclusions.

\section{The anisotropic metric and filed equations}
\label{sm}

Theoretical arguments and indications from the recent observational data support that the existence of anisotropy at early times is a natural phenomenon to investigate. Therefore, it makes sense to consider models of a universe with an initially anisotropic background. The anisotropic Bianchi models may provide an adequate description of anisotropic phase in the history of the universe. The simplest models of the anisotropic universe are the BI model, whose spatial sections are flat, but the scale factors are different in each direction.

The metric for a flat, anisotropic BI cosmological model is given by
\begin{eqnarray}
  &ds^2=dt^2 -A^2(t)dx^2 -B^2(t)dy^2 -C^2(t)dz^2,
  \label{emt}
\end{eqnarray}
where the scale factors $A(t)$, $B(t)$ and $C(t)$ are functions of time only. The non-trivial Christoffel symbols corresponding to this metric are
\begin{eqnarray}
  &\Gamma_{10}^1 = \frac{\dot{A}}{A}, \quad \Gamma_{20}^2 = \frac{\dot{B}}{B}, \quad \Gamma_{30}^3 = \frac{\dot{C}}{C},\nonumber \\
  &\Gamma_{11}^0 = A \dot{A}, \quad \Gamma_{22}^0 = B \dot{B}, \quad \Gamma_{33}^0 = C \dot{C}. \label{eGamma}
\end{eqnarray}
Here, the overhead dot on the scale factors denote differentiation with respect to time $t$, i.e., $d/dt$. Then, with the help of Eq. (\ref{eGamma}), one can get the Ricci scalar of the BI universe as
\begin{eqnarray}
  &R = -2( \frac{\ddot{A}}{A} +\frac{\ddot{B}}{B} +\frac{\ddot{C}}{C} + \frac{\dot{A}}{A} \frac{\dot{B}}{B} + \frac{\dot{B}}{B} \frac{\dot{C}}{C} + \frac{\dot{C}}{C} \frac{\dot{A}}{A} ). \label{eRicci}
\end{eqnarray}

As is well known, the Einstein field equation is given by
\begin{eqnarray}
  &R_{\mu\nu} -\frac{1}{2} g_{\mu\nu} R = T_{\mu\nu},  \label{eEin}
\end{eqnarray}
where $R_{\mu\nu}$ is the Ricci tensor and $T_{\mu\nu}$ is the total energy momentum tensor. The energy momentum tensor of the source with anisotropic pressures along different spatial directions has the form:
\begin{eqnarray}
  &T_{\mu\nu} &= \text{diag} [\rho, -p_x, -p_y, -p_z] \nonumber\\
  &&= \text{diag} [1, -w_x, -w_y, -w_z]\rho, \label{eTm}
\end{eqnarray}
where $\rho$ and $p_i$ are the total energy density and pressure on different directions of the universe, the subscript $i = x$, $y$, $z$ respectively denote the coordinates $x$, $y$ and $z$, the pressure of the universe is related to the energy density through the relation $p_i = w_i \rho$, and $w_i$ are state parameters along different directions.
In our article, we suppose that the universe is filled with DM and DE, the pressure of DM is zero and the pressure of DE is diverse on different directions. Then the energy momentum tensor can be written as
\begin{eqnarray}
  &T_{\mu\nu}^{(m)} = \text{diag} [1,0,0,0] \rho_{m}, \label{epmm}\\
  &T_{\mu\nu}^{(h)} = \text{diag} [1,-(w_{h}+\delta_x),-(w_{h}+\delta_y),-(w_{h}+\delta_z)] \rho_{h}, \label{epxm}
\end{eqnarray}
where $T_{\mu\nu}^{(m)}$ and $T_{\mu\nu}^{(h)}$ are the energy momentum tensor of DM and DE, $\delta_i$ ($i=x,y,z$) are the deviations from the general state parameter (i.e., $w_h$) on different axes (here, $w_h$ is not the mean parameter of the state parameters on different directions, but a general parameter that applies the conservation equation (\ref{erx})),
$\rho_m$ and $\rho_h$ are energy densities of DM and DE, respectively.

The Einstein field equations (i.e., Eq. (\ref{eEin})) for the BI universe now can be written in the following form
\begin{eqnarray}
  &\frac{\ddot{B}}{B} +\frac{\ddot{C}}{C} +\frac{\dot{B}\dot{C}}{BC} = - (w_{h}+\delta_x) \rho_{h}, \label{eE1}\\
  &\frac{\ddot{C}}{C} +\frac{\ddot{A}}{A} +\frac{\dot{C}\dot{A}}{CA} = - (w_{h}+\delta_y) \rho_{h}, \label{eE2}\\
  &\frac{\ddot{A}}{A} +\frac{\ddot{B}}{B} +\frac{\dot{A}\dot{B}}{AB} = - (w_{h}+\delta_z) \rho_{h}, \label{eE3}\\
  &\frac{\dot{A}\dot{B}}{AB} +\frac{\dot{B}\dot{C}}{BC} +\frac{\dot{C}\dot{A}}{CA} = \rho_{m}+\rho_{h}.  \label{eE0}
\end{eqnarray}
Using Eqs. (\ref{eE1})-(\ref{eE0}), the law of energy conservation equation can be written as
\begin{eqnarray}
  &\dot{\rho}_m +3H\rho_{m} +\dot{\rho}_h +3H(1+w_{h})\rho_{h} +(\delta_x H_x +\delta_y H_y +\delta_z H_z) \rho_{h} =0, \label{ec}
\end{eqnarray}
where $H$ is the generalized mean Hubble's parameter and it is defined as
\begin{eqnarray}
  &H = \frac{\dot{a}}{a} = \frac{1}{3} (H_x +H_y +H_z), \label{emH}
\end{eqnarray}
where $a$ is the mean scale factor and $H_x = \frac{\dot{A}}{A}$, $H_y = \frac{\dot{B}}{B}$, $H_z = \frac{\dot{C}}{C}$ are the directional Hubble parameters along $x$, $y$ and $z$ axes, respectively. Using the definitions, the Ricci scalar $R$ can be written in the following form
\begin{eqnarray}
  &R = -6 \big[\dot{H} + (2 +\frac{\Delta}{2}) H^2 \big], \label{eR}
\end{eqnarray}
where $\Delta = \frac{1}{3} \Sigma \bigl( \frac{H_i -H}{H} \bigr)^2$ is a measure of deviation from isotropic expansion. When $\Delta =0$, one can get the isotropic behavior of the model. In the present paper, we used the fact that $B(t) = C(t)$ and $A(t) = C^{m}(t)$. Then, the mean Hubble's parameter reduced to $H = \frac{m+2}{3} H_z$ and the average anisotropy parameter $\Delta = 2(\frac{m-1}{2+m})^2$ (or $m = \frac{2(1+\Delta) \pm 3\sqrt{2\Delta}}{2-\Delta}$) becomes a constant. For $m=1$, the model reduces to isotropic and $\Delta =0$. In the following calculations, we use $m = \frac{2(1+\Delta) + 3\sqrt{2\Delta}}{2-\Delta}$, then according to the work of Companelli et al.\cite{Campanelli2011}, the present value of the average anisotropic parameter is $\Delta = 10^{-5}$ which corresponds to $m = 1.00672$.

As one of the DE models, the HDE may provide more natural solutions to both DE problems at the same time \cite{Samanta2014}. When taking the future event horizon as the choice for the IR cutoff in the HDE, it will raise the causality problem, then the modification of IR cutoff is needed \cite{Cai2007}. Gao et al. \cite{Gao2009} raised the Ricci DE, a new form of HDE whose IR cutoff length corresponding to the Ricci scalar. Granda and Oliveros \cite{Granda2008,Granda2009} also suggested another new type of HDE with modified Ricci scalar (in another words, the IR cutoff is in terms of $H$ and $\dot{H}$).
Here, we would like to take a general holographic Ricci DE model (GHRDE) where the energy density is given by
\begin{eqnarray}
  &\rho_h = 9\frac{1+2m}{(m+2)^2} (\alpha H^2 + 2\beta \dot{H}), \label{erh}
\end{eqnarray}
where $\alpha$ and $\beta$ are model parameters (this model is just like the model Granda and Oliveros raised). It is easy to find that when $\beta = 0$ the IR cutoff reduces to the Hubble length, and when $\alpha = 4 \beta$ the model reduces to the original Ricci DE model.

\section{Interacting GHRDE model}
\label{sr}

The evolution of the BI universe filled with DM and DE will be investigated in this section. We suppose that there is an energy flow between DM and DE. Then, the law of energy conservation equation, i.e., Eq. (\ref{ec}), can be written as
\begin{eqnarray}
  &\dot{\rho}_m +3H\rho_{m} = Q, \label{erm}\\
  &\dot{\rho}_h +3H(1+w_{h})\rho_{h} = -Q, \label{erx}\\
  &3\frac{m \delta_x +\delta_y + \delta_z}{m+ 2} H \rho_{h} =0, \label{edeltac}
\end{eqnarray}
where $Q$ denotes the interaction between DM and DE. Besides, Eq. (\ref{edeltac}) involves the deviations of DE's EoS. Combining Eqs. (\ref{eE1})-(\ref{eE3}) and (\ref{edeltac}), one can obtain that
\begin{eqnarray}
  &\delta_x = \frac{6(m-1) (3H^2 +\dot{H})}{(2+m)^2 \rho_{h}},\label{ed1}\\
  &\delta_y = \delta_z = - \frac{3m(m-1) (3H^2 +\dot{H})}{(2+m)^2 \rho_{h}},\label{ed2}\\
  &w_{h} = -\frac{3(2m+1) (3H^2 +2\dot{H})}{(2+m)^2 \rho_{h}}\label{ewX}.
\end{eqnarray}
From Eqs. (\ref{ed1}) and (\ref{ed2}), one can find that skewness parameters are depended on the evolution of DE and the Hubble' parameter.

Assuming that $Q$ takes the form $Q= 3 b H \rho_{m}$ in the present research. Using Eq. (\ref{erm}), one can obtain
\begin{eqnarray}
  &\rho_{m} = 9\frac{1+2m}{(m+2)^2} H_0^2 \Omega_{m0}\cdot a^{3(b-1)}, \label{erhomx}
\end{eqnarray}
where $H_0$ is the present value of Hubble parameter, $\Omega_{m0} = \rho_{m0} /[9\frac{1+2m}{(m+2)^2} H_0^2]$ is the present value of the dimensionless energy density of DM, and the mean scale factor is defined as $a = C^{(2+m)/3}$, meanwhile, we taking the present value of the scale factor as $a_0 =1$.

Now, let's define that
\begin{eqnarray}
  &h = \frac{H}{H_0}, \quad \tilde{\rho}_m = \frac{(m+2)^2 \rho_m}{9(1+2m) H_0^2}, \quad \tilde{\rho}_h = \frac{(m+2)^2 \rho_h}{9(1+2m) H_0^2}.
\end{eqnarray}
Then, Eqs. (\ref{eE0}), (\ref{erh}) and (\ref{erhomx}) could be rewritten as
\begin{eqnarray}
  &h^2 = \tilde{\rho}_m +\tilde{\rho}_h, \label{eh}\\
  &\tilde{\rho}_h = \alpha h^2 +2\beta \dot{h}, \label{erhh}\\
  &\tilde{\rho}_m = \Omega_{m0} a^{3(b-1)}. \label{ermh}
\end{eqnarray}
Using Eqs. (\ref{eh}), (\ref{erhh}) and (\ref{ermh}), one can obtain a differential equation as following
\begin{eqnarray}
  &\beta \frac{d h^2}{d \ln a} + \alpha h^2 + \Omega_{m0} a^{3(b-1)} =0. \label{ediffH}
\end{eqnarray}
The general solution of the above differential equation is
\begin{eqnarray}
  &h^2 = c_0 a^{-\frac{\alpha-1}{\beta}} + \frac{\Omega_{m0}}{ 1-\alpha +3(1-b)\beta } a^{3(b-1)}.  \label{eHx}
\end{eqnarray}
Here, the free constant $c_0$ can be determined by the initial condition: $h^2|_{a=1} = 1$. Therefore, the free constant $c_0$ is
\begin{eqnarray}
  &c_0 = 1 - \frac{\Omega_{m0}}{ 1-\alpha +3(1-b)\beta }.  \label{ec0}
\end{eqnarray}
Combining Eqs. (\ref{erh}) and (\ref{eHx}), the density of DE can be written as
\begin{eqnarray}
  &\tilde{\rho}_{h} = c_0 a^{-\frac{\alpha-1}{\beta}} +\frac{ \alpha +3(b-1)\beta}{1-\alpha +3(1-b)\beta }H_0^2 \Omega_{m0} a^{3(b-1)}. \label{erhoxx}
\end{eqnarray}

It is easy to find that there are two free parameters in the DE model, in order to reduce the numbers of free parameters we would like to take the present values of $\Omega_{m0}$, $H_0$ and $w_{h0}$ as the boundary conditions. Using Eqs. (\ref{ec}) and (\ref{eh}), we have
\begin{eqnarray}
  &\frac{d h^2}{d\ln a} \big{|}_{a=1} = -3[\Omega_{m0}+(1+w_{h0})(1-\Omega_{m0})],
\end{eqnarray}
meanwhile, from Eq. (\ref{eHx}) one can get that
\begin{eqnarray}
  &\frac{d h^2}{d\ln a} \big{|}_{a=1} = \frac{1-\alpha -\Omega_{m0}}{\beta}.
\end{eqnarray}
Combine these two solutions, we obtain that
\begin{eqnarray}
  &\alpha = 3\beta +(1+3\beta w_{h0})(1-\Omega_{m0}). \label{eal}
\end{eqnarray}
Thus, if we take the present observation data into Eq. (\ref{eal}), the free parameter $\alpha$ can be determined by the value of $\beta$. In our paper,  the present value of GHRDE's state parameter is chosen as $w_{h0} = -1$.
\begin{figure}
 \centering
  \includegraphics[scale=.73]{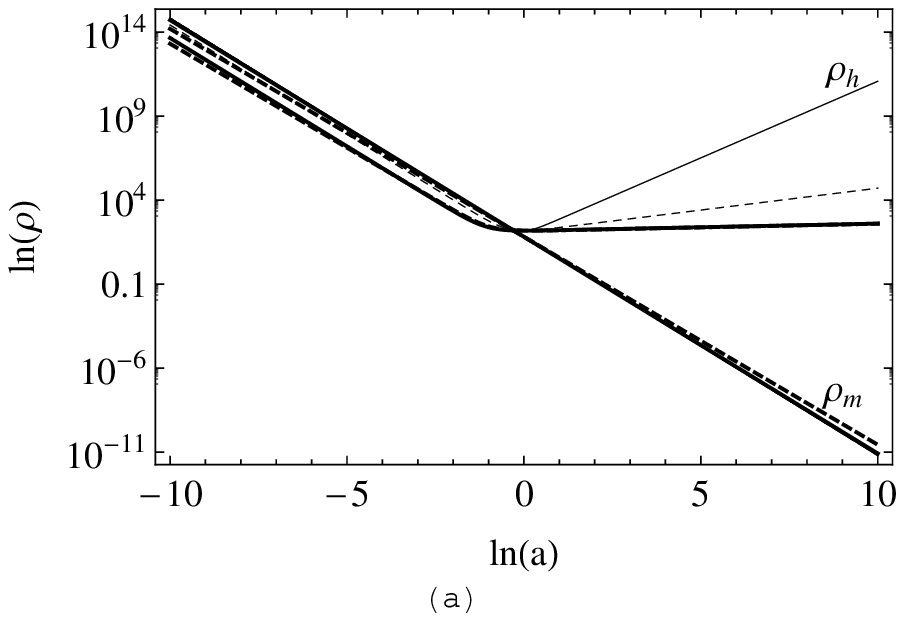}
  \includegraphics[scale=.7]{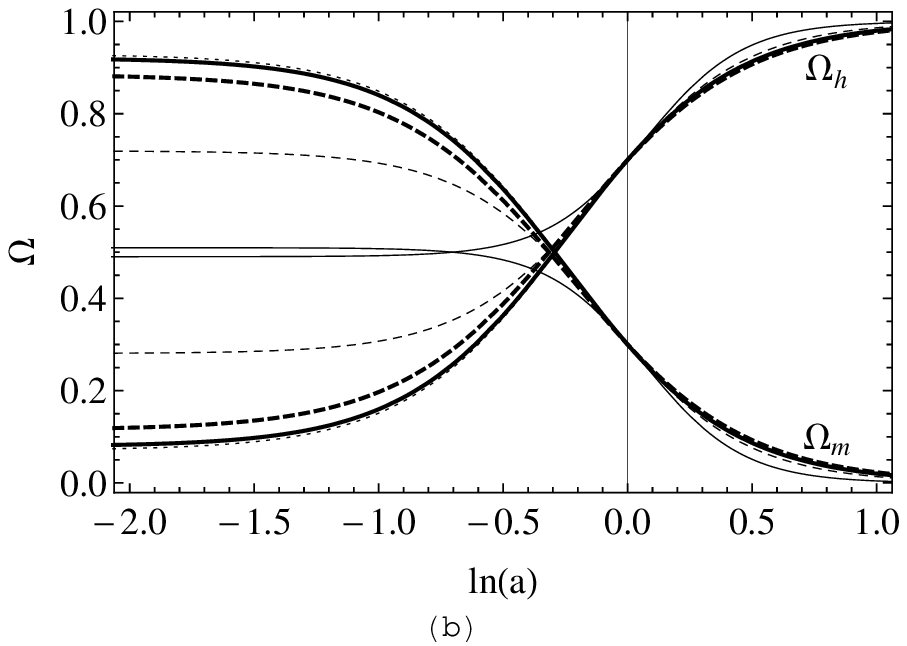}
  \caption{The graphics show the variation of energy densities of DM and DE against $\ln{a}$ under the condition: $\Delta = 10^{-5}$, $H_0^2 = 72$, and $\Omega_{m0} = 0.3$. Here, the solid, dashed, dotted, thick solid and thick dashed lines are for $\beta = 0.1$, $b= 0.001$; $\beta = 0.2$, $b= 0.001$; $\beta = 0.3$, $b= 0.001$; $\beta = 0.3$, $b= 0.010$; $\beta = 0.3$, $b= 0.050$, respectively.}
  \label{fn1}
\end{figure}
\begin{figure}
 \centering
  \includegraphics[scale=.7]{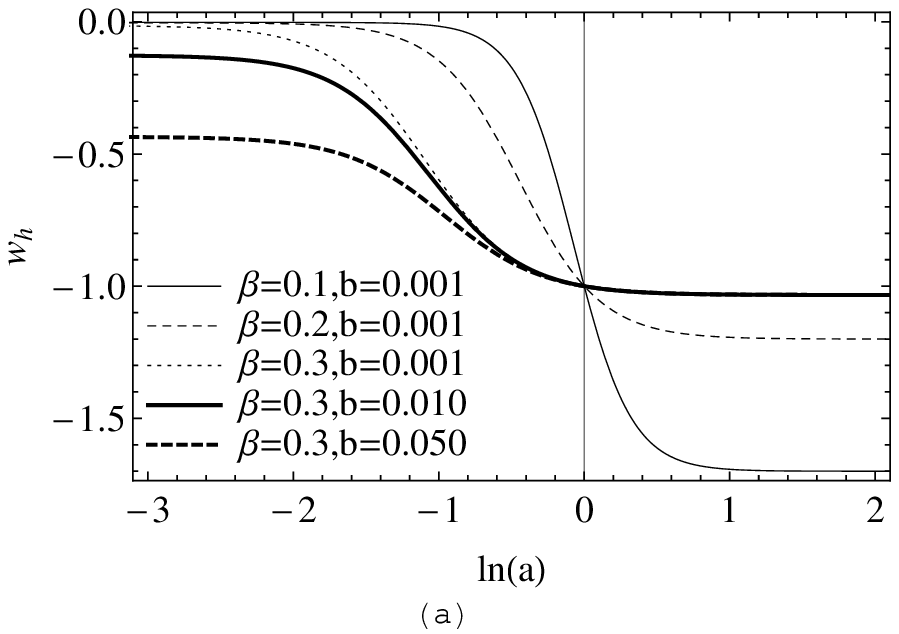}
  \includegraphics[scale=.7]{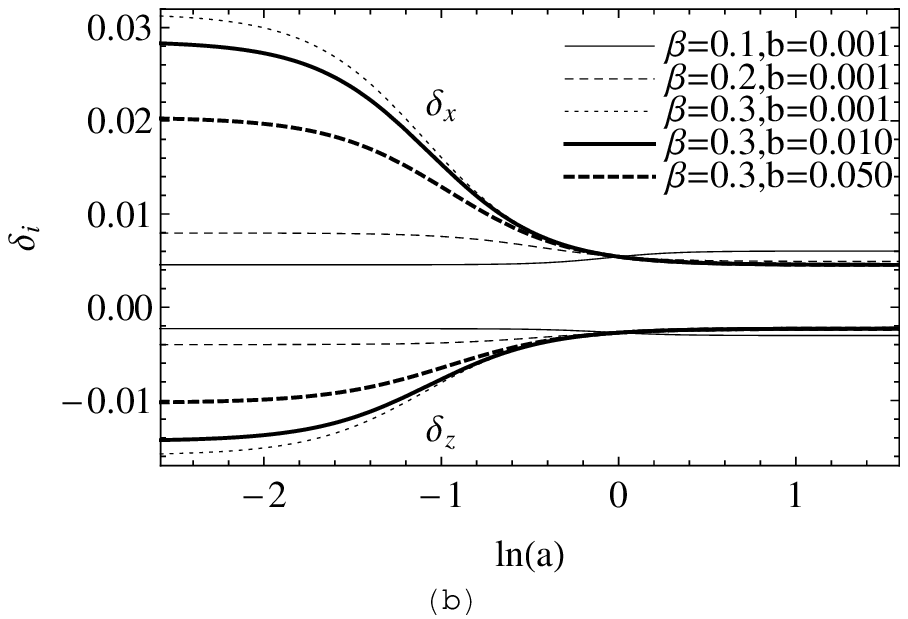}
  \caption{The first graphic is for the DE's EoS with respect to $\ln a$, the second graphic is for the deviations of state parameter with respect to $\ln a$. Here we choose $\Delta = 10^{-5}$, $H_0^2 = 72$, and $\Omega_{m0} = 0.3$.}
  \label{fn2}
\end{figure}
\begin{figure}
	\centering
	\includegraphics[scale=.5]{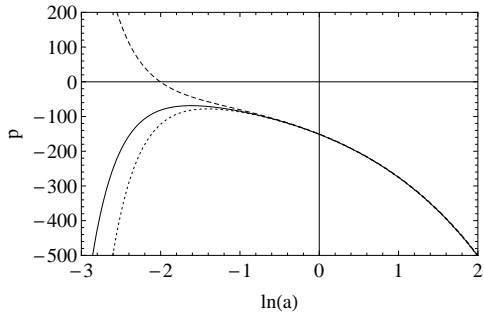}
	\caption{The universe's pressure $p$ with respect to $\ln a$. The solid line is for the general pressure $p = w_h \rho_h$, the dashed line is for the pressure along $x$ axis and the dotted line is for the pressure along $y$ and $z$ axes. Here we choose $\Delta = 10^{-5}$, $\beta = 0.2$, $b=0.001$, $H_0^2 = 72$, and $\Omega_{m0} = 0.3$.}
	\label{fn3}
\end{figure}

In Fig. \ref{fn1}, we plotted the energy densities of DM and GHRDE as a function of the mean scale factor. We find that in the early time of the universe, the energy density of GHRDE is comparable with that of DM, but in the future it's DE dominating. We can also find that the model parameter $\beta$ has a great influence on the energy density of GHRDE in the future. According to Fig. \ref{fn2}(a), one can find that with the value of $\beta$ increasing, the future value of $w_h$ decreases, moreover, as the interacting constant $b$ increasing, the value of $w_h$ in the early time decreasing. The behavior of $w_h$ indicates that GHRDE is matter-like in the past and phantom-like in the future. From Fig. \ref{fn2}(b), it is easy to find that the universe is high anisotropic in the early time and more homogeneous in the future. Combining Figs. \ref{fn2} and \ref{fn3}, we find that the pressure of the universe varies greatly in different directions in the early time and it becomes isotropic in the future. Above all, the model we use can explain the high anisotropic of the early universe and homogeneous from near past to the future.

\section{The Generalized second law of gravitational thermodynamics}
\label{st}

The GSLT in the anisotropic BI universe driven by the GHRDE would be shown in this section. The horizon surface would be taken as a boundary of the thermodynamic system. Using the Gibbs' law of thermodynamics, the entropy of the BI universe with DE and interacting DM inside the horizon is given by
\begin{eqnarray}
  &T_k d S_k = d E_k + p_k d V, \label{eGibb}
\end{eqnarray}
where $T_k$, $S_k$, $E_k$, $p_k$, and $V$ are the temperature, entropy, energy ($E_k = \rho_k V$), pressure, and volume, respectively, and the subscript $k=m$, $h$ represent DM and GHRDE, respectively.  Here, the volume containing DM and GHRDE with the radius of horizon $R_H$ is $V = \frac{4}{3} \pi R_H^3$.
To examine the change rate of the total entropy, one should give a horizon. We would like to check the validity of the GSLT on the horizon of the general Ricci scalar radius $R_{gr} = [\alpha H^2 +2\beta \dot{H} ]^{-1/2}$. The Hawking temperature and the entropy of the horizon are defined as $T_H = 1/(2\pi R_{gr})$ and $S_H = \kappa \pi R_{gr}^2$.

In view of Gibbs' equation and the condition that $dS$ be a differential, which is given by $\frac{\partial^2 S_k}{\partial V \partial T_k} = \frac{\partial^2 S_k}{\partial T_k \partial V}$. This leads to the relation that
\begin{eqnarray}
    &dp_k = \frac{\rho_k + p_k}{T_k} dT_k. \label{eTk}
\end{eqnarray}
Using Eq. (\ref{ec}), this relation can be rewritten as $d\ln T_k/d\ln a = -3 \partial p_k/ \partial \rho_k = -3w_k$ \cite{Weinberg1971,Pavon2009,Radicella2014}.
In this case, the temperature of dark fluid inside the horizon will be
\begin{eqnarray}
  & T_k = T_{k0} \text{exp} \left[ -3 \int_0^{\ln a} w_k(x) dx \right]. \label{eTf}
\end{eqnarray}
From this equation, one know that the temperature of the dark fluid will change with time going and will never be negative.
Plugging Eq. (\ref{eTk}) into Eq. (\ref{eGibb}), the Gibbs' equation will be rewritten as
\begin{eqnarray}
    &dS_k = \frac{d[(\rho_k +p_k) V]}{T_k} -(\rho_k +p_k) V \frac{dT_k}{T_k^2} = d \left[ \frac{(\rho_k + p_k) V}{T_k} + const \right],
\end{eqnarray}
where $const$ is a constant. Hence the entropy can be defined as
\begin{eqnarray}
    & S_k = \frac{\rho_k + p_k}{ T_k } V = \frac{1+w_k}{T_K} \rho_k V. \label{eS}
\end{eqnarray}
Eq. (\ref{eS}) shows that the positive or negative of the entropy $S_k$ is depending on the EoS of DM and GHRDE. And it is easy to find that when $w_h < -1$, $S_h <0$, which indicates that when the GHRDE is phantom-like, its entropy's change rate will be negative \cite{Radicella2010B,Radicella2012}. According to Fig. \ref{fn2}, GHRDE is phantom-like from now to the future, which shows that the GSLT will be invalid on GHRDE from now to the future.

In consideration of that there is interaction between DM and DE, without loss of generality, we suppose that the temperature of DM and GHRDE are equal to each other. Then the change rate of the internal entropy $S_I$, which is defined as $S_I = S_m +S_h$, could be written as
\begin{eqnarray}
  &S_{I,eff}' = 4 \pi R_{gr}^2 (1+w_{eff}) (R_{gr}H'-R_{gr}) \frac{\rho}{T_{eff}}, \label{eSeff}
\end{eqnarray}
Using Eq. (\ref{eTf}), the temperature can be obtained as
\begin{eqnarray}
&T_{eff} \propto h^2 a^3.
\end{eqnarray}	

\begin{figure}
  \centering
  \includegraphics[scale=.7]{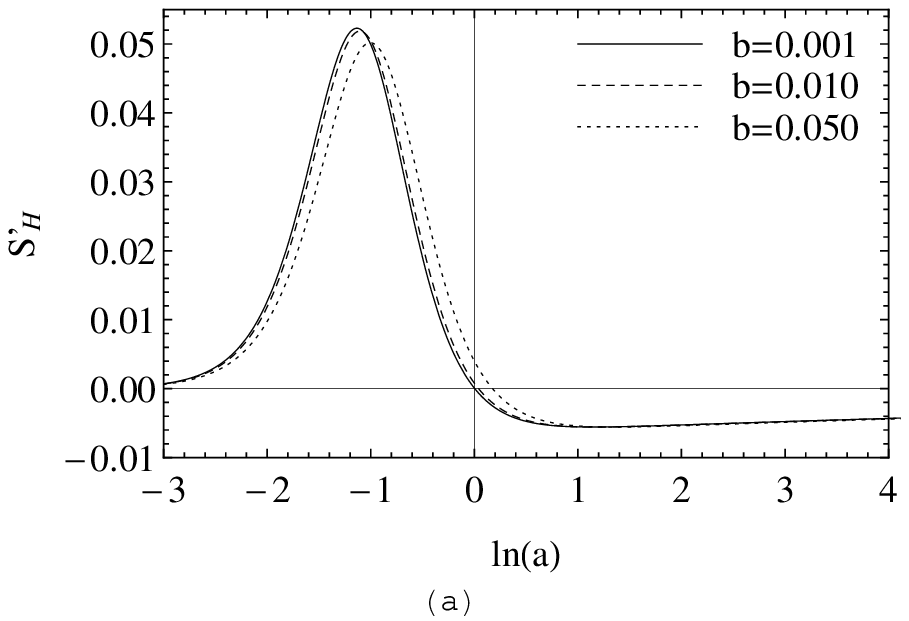}
  \includegraphics[scale=.68]{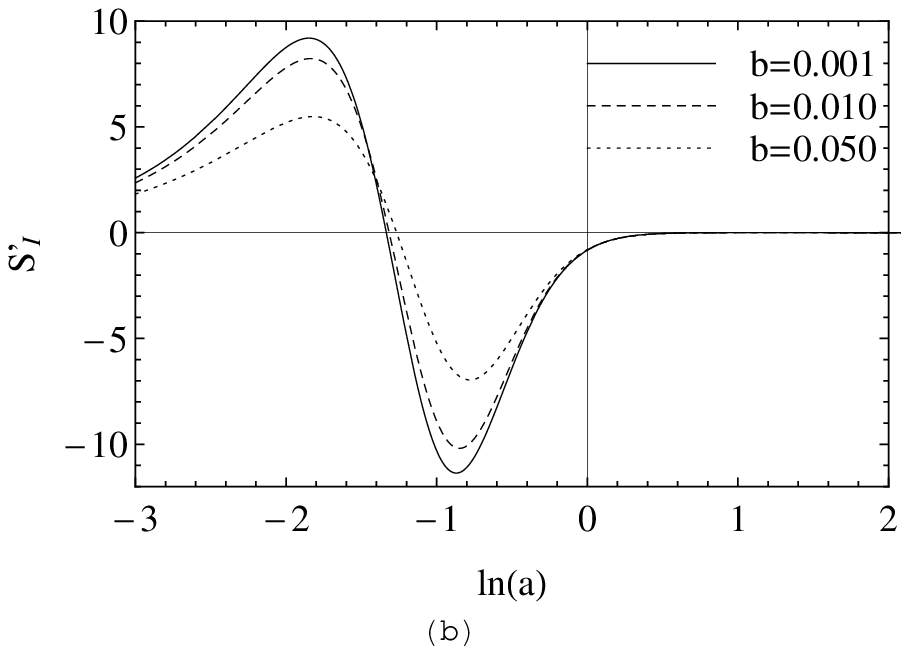}\\
  \includegraphics[scale=.7]{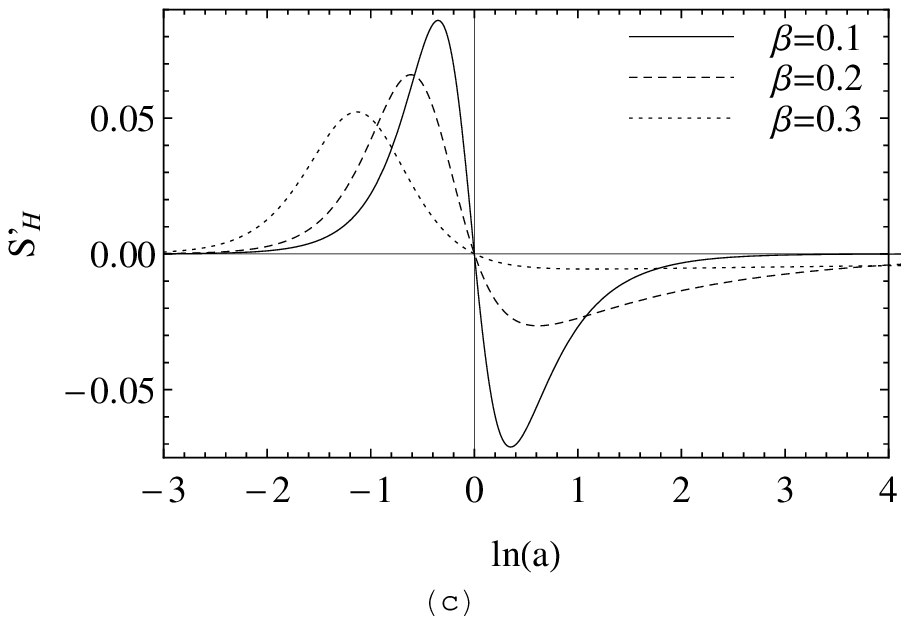}
  \includegraphics[scale=.68]{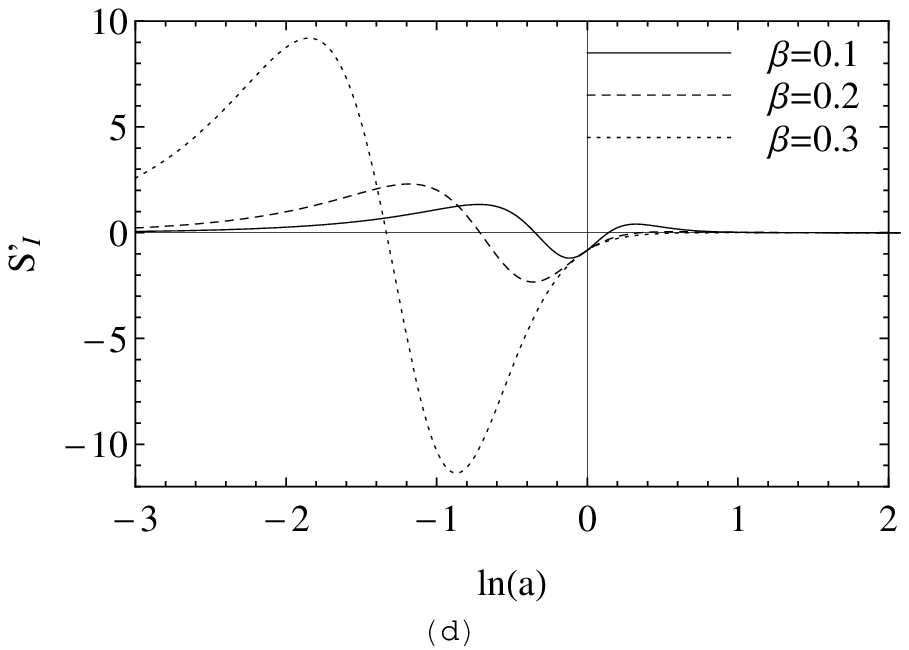}
  \caption{The variation of $S'_{gr}k$ and $S'_I$ with respect to $\ln a$ under different cases. Here we choose $\Delta = 10^{-5}$, $H_0^2 = 72$ and $\Omega_{m0} = 0.3$. For the two graphics ahead, we choose $\beta = 0.3$, and for the last two graphics, we use $b=0.001$.}
  \label{fn5}
\end{figure}

The variation of $S'_{gr}$ and $S'_{I}$ against $\ln a$ is shown in Fig. \ref{fn5}. From Figs. \ref{fn5} (a) and \ref{fn5} (b), one can find that the strength of interaction has little effect on the change rate of entropy. But, combining Figs. \ref{fn5} (c) and \ref{fn5} (d), it shows that the model parameter has a major impact on the change rate of entropy, as $\beta$ increases, the negative $S_I'$ occurs earlier.

From Fig. \ref{fn5}, it is easy to find that for the change rate of entropy on the horizon, it is negative from now to the future, and for the entropy of the internal, the negative change rate occurs from the near past to the near future. As we know that $S' < 0$ indicates the entropy decreases and the GSLT became invalid. This phenomenon casts doubts on the soundness of the GHRDE model. On the other hand, considering that the HDE has gained many success in explaining the evolution of the universe, we surmise that the phenomenon may be caused by the galaxies formation. When the galaxies come into being, part of the universe goes from ``order'' to ``disorder'', the entropy is increasing, not decreasing. Because the universe's total entropy is not decreasing, so this may be helpful in balancing the entropy decreasing causes by DE. Furthermore, contrast $S_H'$ and $S_I'$ in Fig. \ref{fn5}, we find that the negative $S_I'$ occurs more early than that of $S_H'$, this may indicates that the influences of galaxies formation passes from the internal universe to the horizon.

\section{Conclusions}
\label{sc}

The BI universe filled with DM and GHRDE is studied in the present paper. In this paper, a GHRDE model is taken into account, whose IR cutoff radius is taken as $L^2 \propto (\alpha H^2 +2\beta \dot{H})^{-1}$. This model can be reduced to the original Ricci DE model when $\alpha/ \beta = 4$. By considering the energy densities, we find that the energy density of GHRDE is comparable with that of DM in the early time and GHRDE dominating in the future. Through analysing the behavior of state parameter of GHRDE, we obtain that the GHRDE transits from matter-like to phantom-like through the history. The behavior of the deviations of state parameter and the pressure of the universe indicate that the universe is high anisotropic in the early time, but more homogeneous from near past to the future.

The GSLT of BI universe containing DM and DE is studied in Sect. \ref{st}.
The temperature of DM and GHRDE can not be equal to that of the horizon, and considering there is an interaction between DM and GHRDE. The temperature of the dark fluid can be determined by $d\ln T_{eff} /d\ln a = -3 \partial p/ \partial \rho$. In this case, we find that the entropy of GHRDE is negative when it is phantom-like.
Besides, it turns out that when taking the general Ricci scalar radius $R_{gr}$ as the horizon, the curves of $S'$ have a positive maximum in the early time and a negative minimum in the near past or the future, which indicates that the GSLT is valid in the early universe but invalid from near past to the near future.
Then we surmise that the formation of the galaxies may be helpful in explaining this phenomenon, because when the galaxies taking shape, part of the universe goes from ``order'' to ``disorder'', the total galaxies' entropy is increasing, which may be helpful in explaining the universe's total entropy is not decreasing. And the negative $S'$ occurs in different periods indicates that the influences of galaxies formation could wipe from the internal universe to the horizon.

\section*{Acknowledgements}

This work was supported in part by the National Natural Science Foundation of China (Grants No. 11565016, No. 11405076), the Science Foundation of the Education Department of Yunnan Province (Grant No. 2014Y066), and the Talent Cultivation Foundation of Kunming University of Science and Technology (Grants No. KKSY201207053, No. KKSY201356060). Yu Zhang would like to acknowledge the support of the working Funds of the Introduced High-level Talents of Yunnan Province from the Department of Human Resources and Social Security of Yunnan Province.

\end{document}